# On the Critical Exponents for the λ Transition in Liquid Helium

### A. A. Pogorelov and I. M. Suslov

*Kapitza Institute for Physical Problems, Russian Academy of Sciences, ul. Kosygina 2, Moscow, 119334 Russia*
*e-mail: suslov@kapitza.ras.ru*

The use of a new method for summing divergent series makes it possible to significantly increase the accuracy of determining the critical exponents from the field theoretical renormalization group. The exponent value ν = 0.6700 ± 0.0006 for the λ transition in liquid helium is in good agreement with the experiment, but contradicts the last theoretical results based on using high-temperature series, the Monte Carlo method, and their synthesis.

The critical exponents for the λ transition in liquid $^4$He are the most accurately measured exponents. The measurement of the density of the superfluid component $\rho_s \sim \tau^\zeta$ ($\tau$ is the distance to the transition) in the experiments on the second sound made it possible to determine [in view of the relation $\zeta = (d-2)\nu$, where $d$ is the space dimension] the critical exponent ν of the correlation radius as

$$\nu = 0.6705 \pm 0.0006 \quad [1]. \tag{1}$$

The measurements [2–4] of the specific-heat exponent α, which are performed on satellites in order to avoid the gravity-induced broadening of the transition (1-cm helium column pressure shifts the transition by $10^{-6}$ K), are more accurate:

$$\alpha = -0.01285 \pm 0.00038 \quad [2],$$
$$\alpha = -0.01056 \pm 0.00038 \quad [3], \tag{2}$$
$$\alpha = -0.0127 \pm 0.0003 \quad [4].$$

The differences in the values are attributed to the ambiguity of the interpretation. Results (1) and (2) can be compared to each other via the scaling relation α = 2 − $d\nu$ (Fig. 1). They are in good agreement with the predictions ν = 0.669 ± 0.003 [5] and ν = 0.6695 ± 0.0020 [6] of the field theoretical renormalization group approach, which were made about 30 years ago and remained of record accuracy for a long time (recent improvement provides ν = 0.6703 ± 0.0015 [7]). More close agreement with the experimental values was achieved in the variational perturbation theory [8–11] based on the same information. The intrigue of the last years is that the more accurate theoretical predictions based on the use of high-temperature series [12], the Monte Carlo method [13, 14], and their synthesis [15, 16] are concentrated at higher ν values and begin to contradict the experimental results (Fig. 1). Opinions that the experimental results are unsatisfactory and further investigations are necessary appear [16].

Campostrini et al. [16] stated that the accuracy of the field theoretical approach cannot favor the experimental results [1–4] or new theoretical results [14–16]. The aim of this work is to contest this statement: the use of a new algorithm for summing divergent series [17–19] makes it possible to refine the predictions of the field theoretical renormalization group approach and to certainly resolve the contradiction in favor of the experimental values (see Fig. 1).

The initial information is the first seven coefficients of the expansion of the renormalization group functions $\beta(g)$, $\eta(g)$, and $\eta_2(g)$ [5, 7] and their high-order asymptotic expressions [20] calculated by the Lipatov method [21],

$$\beta(g) = -g + g^2 - 0.402962963g^3 + 0.314916942g^4$$
$$- 0.31792848g^5 + 0.3911025g^6 - 0.552449g^7$$
$$+ \ldots + ca^N \Gamma(N+b)g^N + \ldots,$$

$$\eta(g) = (8/675)g^2 + 0.0009873600g^3$$
$$+ 0.0018368107g^4 - 0.0005863264g^5$$
$$+ 0.0012513930g^6 - 0.001395129g^7 \tag{3}$$
$$+ \ldots + c'a^N \Gamma(N+b')g^N + \ldots,$$

$$\eta_2(g) = -(2/5)g + (2/25)g^2 - 0.0495134446g^3$$
$$+ 0.0407881055g^4 - 0.0437619509g^5$$
$$+ 0.0555575703g^6 - 0.08041336g^7$$
$$+ \ldots + c''a^N \Gamma(N+b)g^N + \ldots,$$

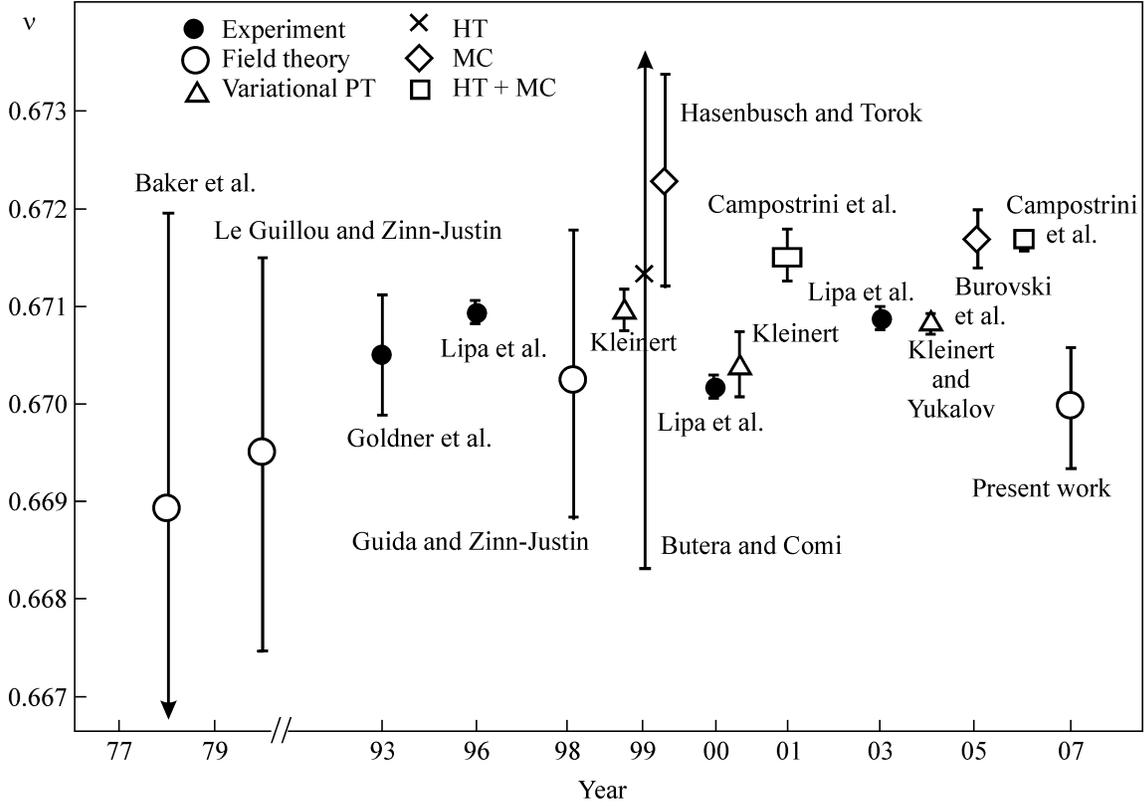

**Fig. 1.** Experimental and theoretical results for the exponent ν.

where

$$a = 0.132996798, \quad b = b' + 1 = 5,$$
$$c = 0.016302, \quad c' = 0.0008798, \quad c'' = 0.0030836. \quad (4)$$

According to the field theoretical renormalization group approach, it is necessary to determine a nontrivial root $g*$ of the equation $\beta(g) = 0$ that specifies the stationary point; after that, the critical exponents $\eta$ and $\nu$, as well as the exponent $\omega$ of the correction to scaling, are given by the expressions

$$\eta = \eta(g*), \quad \nu^{-1} = 2 - \eta(g*) + \eta_2(g*),$$
$$\omega = \beta'(g*). \quad (5)$$

The summation procedure is based on the fact that the series

$$W(g) = \sum_{N = N_0}^{\infty} W_N (-g)^N \quad (6)$$

whose coefficients have the asymptotic behavior $W_N^{as} = ca^N \Gamma(N + b)$, after the Borel transform

$$W(g) = \int_0^{\infty} dx\, e^{-x} x^{b_0 - 1} B(gx),$$
$$\qquad (7)$$
$$B(z) = \sum_{N = N_0}^{\infty} B_N (-z)^N, \quad B_N = \frac{W_N}{\Gamma(N + b_0)},$$

where $b_0$ is an arbitrary parameter and the conformal transformation $z = u/(1 - u)a^{1/}$, reduces to a convergent power series in $u$ with the coefficients

$$U_0 = B_0, \quad U_N = \sum_{K = 1}^{N} \frac{B_K}{a^K} (-1)^K C_{N-1}^{K-1} \quad (N \geq 1), \quad (8)$$

whose asymptotic behavior at $N \longrightarrow \infty$,

$$U_N = U_\infty N^{\alpha - 1}, \quad U_\infty = \frac{W_\infty}{a^\alpha \Gamma(\alpha) \Gamma(b_0 + \alpha)} \quad (9)$$

determines the strong-coupling asymptotic expression for the function $W(g)$,

$$W(g) = W_\infty g^\alpha \quad (g \longrightarrow \infty). \quad (10)$$

---

[1/] This conformal transformation differs from that used in [6, 7]; its advantage is that the increase of random errors in coefficients $U_N$ (8) is much slower and the algorithm is stable with respect to smooth errors [17].

The coefficients $U_N$ for $N \lesssim 40$ are calculated by Eq. (8) and then are continued according to power law (9) in order to avoid the catastrophic increase in errors [17]. Thus, all the coefficients of the convergent series are known and this series can be summed with an arbitrary accuracy. This completely removes the problem of the dependence of the results on variation in the summation procedure, which is the main disadvantage of the commonly accepted methods. Only the dependence on the method for the interpolation of the coefficient functions remains, which has direct physical meaning and is associated with the incompleteness of the initial information. The interpolation is performed for the reduced coefficient function[2]

$$F_N = \frac{W_N}{W_N^{as}} = 1 + \frac{A_1}{N - \tilde{N}} + \ldots + \frac{A_K}{(N - \tilde{N})^K} + \ldots \quad (11)$$

by cutting the series and choosing the coefficients $A_K$ from the correspondence with the known values of the coefficients $W_{L_0}$, $W_{L_0+1}$, $\ldots$, $W_L$. The asymptotic expression is taken in the optimal form $W_N^{as} = ca^N N^{b-1/2} \Gamma(N + 1/2)$ [17], and the parameter $\tilde{N}$ is used to analyze uncertainty in the results. The $L_0$ value sometimes does not coincide with $N_0$ appearing in Eq. (6). Indeed, the coefficient function $W_N$ continued to the complex plane has a singularity at the point $N = \alpha$, where $\alpha$ is the exponent of the strong-coupling asymptotic expression given by Eq. (10) [17]. If the exponent $\alpha$ is larger than $N_0$, the interpolation with the use of all the coefficients is inapplicable: it is necessary to take

$$W(g) = W_{N_0} g^{N_0} + \ldots + W_{N_1} g^{N_1} + \tilde{W}(g),$$
$$N_1 = [\alpha], \quad (12)$$

sum the series for $\tilde{W}(g)$, and add the contribution from the separated terms; thus, the value $[\alpha] + 1$, where $[\ldots]$ stands for the integer part of a number, is taken for $L_0$. Analysis of two-dimensional case [22] shows that $\alpha$ is larger than $N_0$ for almost all the functions.

According to the tradition, we sum the series not only for the functions $\beta(g)$ $\eta(g)$ and $\eta_2(g)$ but also for the functions $\nu^{-1}(g) = 2 + \eta_2(g) - \eta(g)$ and $\gamma^{-1}(g) = 1 - \eta_2(g)/(2 - \eta(g))$ in order to verify the self-consistency of the results. Following [22], we allow the interpolation curves that pass through all the known points, are smooth, do not have significant kinks at noninteger $N$ values, and rapidly achieve the asymptotic behavior at large $N$ values.

**Function $\beta(g)$.** All the interpolations with $L_0 = 1$ are unsatisfactory: the interpolation curves rapidly achieving the asymptotic behavior exhibit a sharp kink in the interval $1 < N < 2$, thereby indicating a singularity in this interval. The estimate of strong-coupling asymptotic expression (see Fig. 2a) yields $\alpha \approx 1$, thereby confirming the singularity at $N \approx 1$ and indicating that the choice $L_0 = 2$ is correct. The interpolation curves with $\tilde{N} < -1.0$ exhibit significant nonmonotonicity at large $N$ values, and the curves with $\tilde{N} > 1.4$ have a kink in the interval $2 < N < 3$ (see Fig. 2b). Thus, the "natural" interpolations correspond to the interval $-1.0 < \tilde{N} < 1.4$. The summation results are shown in Fig. 2b, which indicate that

$$g^* = 1.406{-}1.410, \quad \omega = 0.774{-}0.783. \quad (13)$$

The $g^*$ value is in agreement with the results of early works ($g^* = 1.406 \pm 0.005$ [5] and $g^* = 1.406 \pm 0.004$ [6]) and indicates that the more recent value $g^* = 1.403 \pm 0.003$ obtained in [7] is doubtful.[3]

**Function $\eta(g)$.** According to Eq. (3), the expansion for $\eta(g)$ begins with $g^2$. We fail to obtain satisfactory interpolations with $L_0 = 2$: the curves rapidly approaching the asymptotic behavior exhibit a kink in the interval $2 < N < 3$, thereby indicating that the exponent $\alpha$ lies in the same interval. Indeed, the estimate of the strong-coupling asymptotic expression for $L_0 = 3$ (see Fig. 3a) gives

$$\tilde{\alpha} = 2.00 \pm 0.02, \quad \tilde{W}_\infty = 0.44 \pm 0.06. \quad (14)$$

The satisfactory interpolation curves (see Fig. 3b) exist only for $1.6 < \tilde{N} < 2.3$. They could be considered unsatisfactory because they have a kink for $3 < N < 4$; however, the curves of such a shape provide the exact $\eta$ value in the two-dimensional case [22]. Such interpolations are allowable because the amplitude of oscillations of the coefficient function is on the order of the amplitude of oscillations of the known coefficients. The summation results are shown in the inset in Fig. 3b.

**Functions $\eta_2(g)$, $\nu^{-1}(g)$, and $\gamma^{-1}(g)$.** The interpolation curves with $L_0 = 1$ are inadmissible because they have kinks in the interval $1 < N < 2$. The interpolation curves with $L_0 = 2$ for the function $\nu^{-1}(g)$ are also unacceptable because they exhibit kinks in the interval $2 < N < 3$. Analysis of the strong-coupling asymptotic expression for $L_0 = 2$ provides $\tilde{\alpha} = 0.63 \pm 0.2$ for $\eta_2(g)$ and $\tilde{\alpha} = 0.40 \pm 0.22$ for $\gamma^{-1}(g)$. For the $\nu^{-1}(g)$ function at $L_0 = 3$ (see Fig. 4a), we obtain

$$\tilde{\alpha} = 2.00 \pm 0.02, \quad \tilde{W}_\infty = 0.3 \pm 0.2. \quad (15)$$

---

[2] Corrections to the Lipatov asymptotic expression have the form of the regular expansion in $1/N$ and, after resummation, reduce to form (11).

[3] Note that the results for $g^*$ in [5–7] are based on the same information.

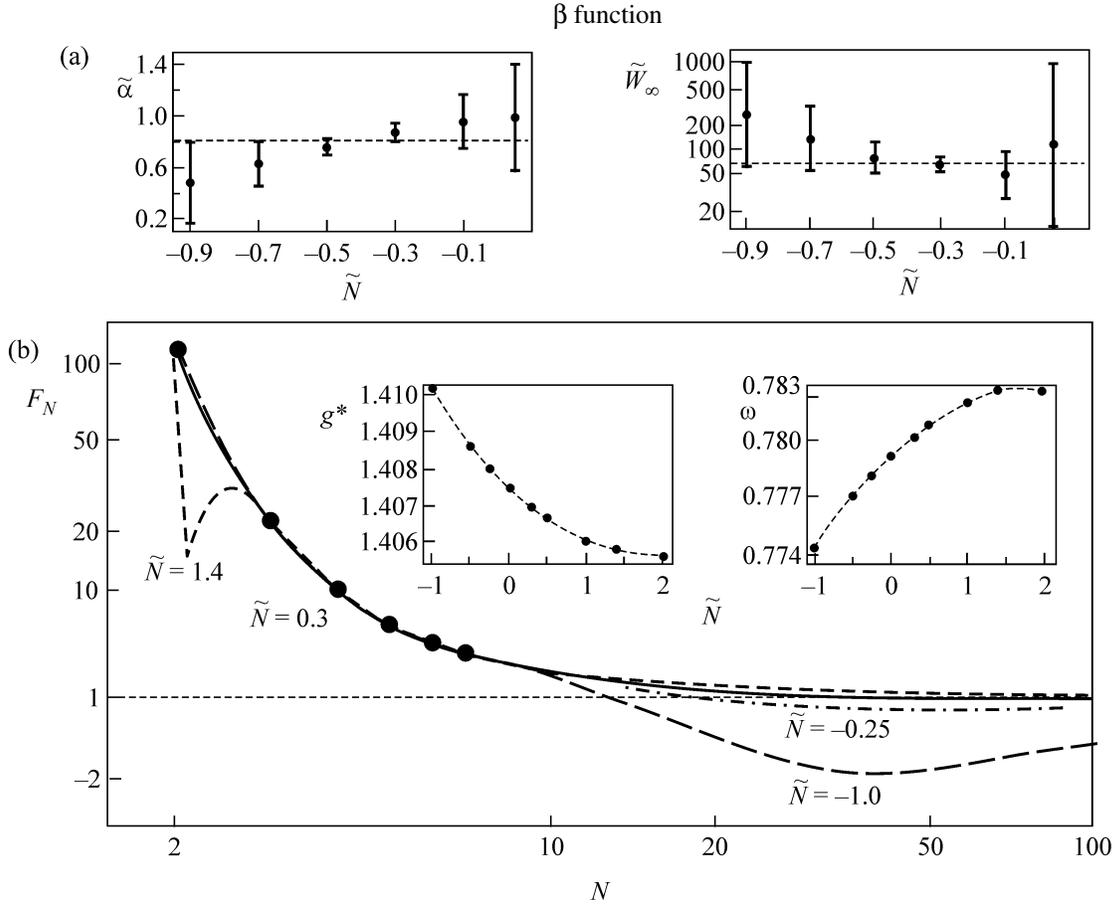

β function

Fig. 2. (a) Parameters of the strong-coupling asymptotic expression for the function $\beta(g)$ (working interval $21 < N < 40$). Uncertainty in the results is determined according to [17], but the error corridor is extended. (b) Interpolation curves for the expansion coefficients of the function $\beta(g)$ and summation results for $g^*$ and $\omega$ (logarithmic scale for $F_N + 5$).

Taking into account the terms separated in Eq. (12), we arrive to results for large $g$ values

$$\nu^{-1}(g) \approx -(0.3 \pm 0.2)g^2, \quad \eta_2(g) \approx -0.4g,$$
$$\gamma^{-1}(g) \approx -0.2g, \quad \eta(g) \approx (0.45 \pm 0.06)g^2. \quad (16)$$

The relation $\nu^{-1}(g) = 2 + \eta_2(g) - \eta(g)$ is satisfied for the asymptotic expressions within the accuracy. The relation $\gamma^{-1}(g) = 1 + \eta_2(g)/(2 - \eta(g))$ is satisfied only when the function $\eta(g)$ is disregarded: the expansion coefficients for this function are small, and so the function is small for $g \lesssim 10$, but its strong-coupling asymptotics varies more rapidly. For this reason, it is expected that the functions $\gamma^{-1}(g)$ and $\eta_2(g)$ first increase linearly to sufficiently large $g$ values (this behavior is looking as the true asymptotic behavior in the approximate analysis) and then their behavior is distorted by the function $\eta(g)$. The function $\gamma^{-1}(g)$ either begins to decrease [if the behavior $\eta_2(g) \sim g$ holds] or approaches a constant [if the asymptotic behavior of the function $\eta_2(g)$ contains a contribution $\sim g^2$]. In any case, the coefficient function for the function $\gamma^{-1}(g)$ is regular for $N \geq 1$, but it contains a smeared singularity at $N \approx 1$. For this reason, the series for the function $\gamma^{-1}(g)$ is summed at $L_0 = 1$, but without the restriction of kinks for noninteger $N$ values. The summation of the series for the functions $\eta_2(g)$ and $\nu^{-1}(g)$ is performed at $L_0 = 3$ in order to take into account a possible singularity at $N \approx 2$.[4] Figures 4b–4d show the allowable interpolations and summation results.

The result

$$\nu = 0.6700 \pm 0.0006 \quad (17)$$

obtained by the direct summation of the series for the function $\nu^{-1}(g)$ is of most interest and can be compared (see Table 1) to the results of the summation of other series[5] in view of the relations $\nu = \gamma(2 - \eta)$, $\nu^{-1} = 2 +$

---

[4] The summation of the series for the function $\eta_2(g)$ at $L_0 = 2$ yields $\eta_2 = -0.4744(7)$, i.e., the same result as in Fig. 4c, but with a smaller error.

[5] According to the $1/n$ expansion, the functions $\nu^{-1}(g)$ and $\gamma^{-1}(g)$ have a root at $g \sim 1$. Owing to the corresponding singularities in the functions $\nu(g)$ and $\gamma(g)$, the series for them have oscillating coefficients and are poorly summarized. For the same reason, we do not try to sum the series for the function $\alpha(g) = 2 - d\nu(g)$.

η-function

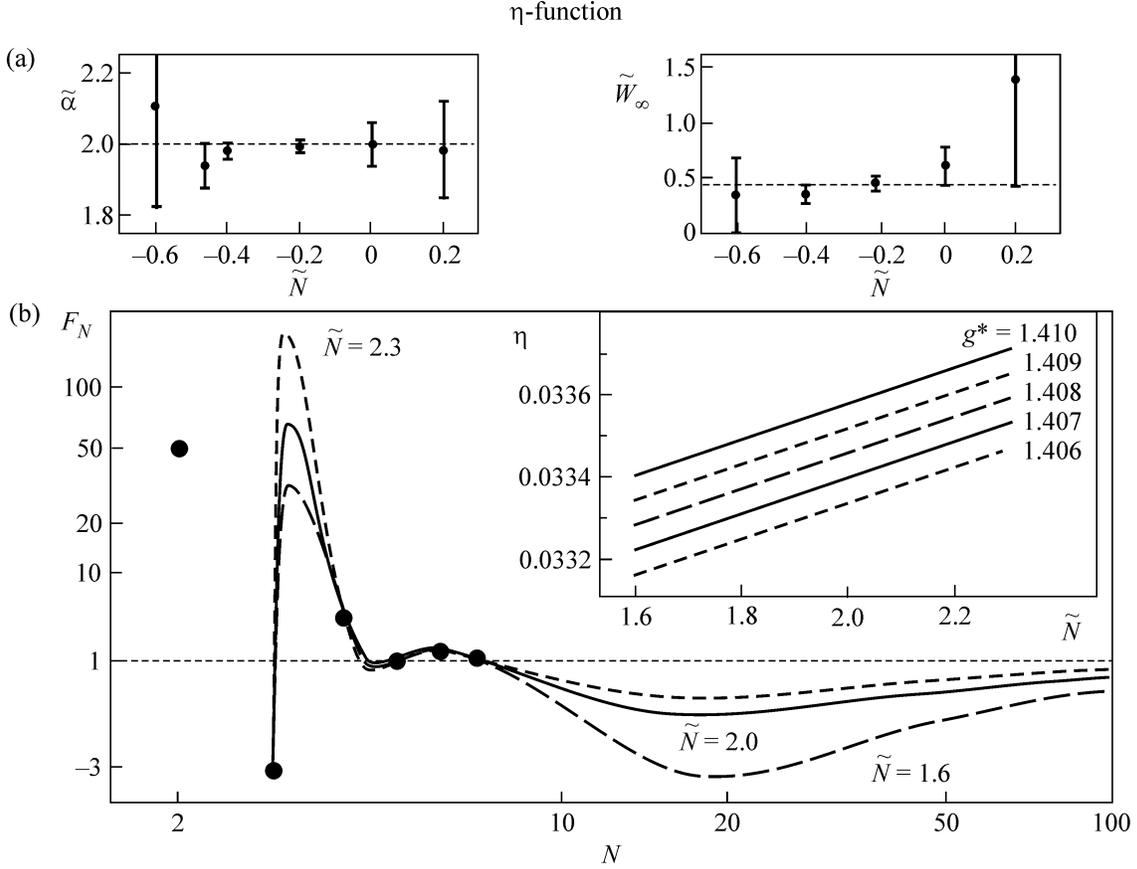

**Fig. 3.** (a) Parameters of the strong-coupling asymptotic expression for the function η(g) (working interval 23 < N < 40). (b) Allowable interpolations for the function η(g) and summation results for g = g*.

$\eta_2 - \eta$ and $\nu = (1 - \gamma)/\eta_2$. According to Table 1, result (17) almost coincides with the union of the second and third estimates (somewhat more accurate) and is contained in the fourth estimate (much less accurate and ignored below). The relative shift of the central values for first three estimates can be treated as the scale of the systematic error,

$$\delta_{syst} \approx 0.0002, \qquad (18)$$

which appears because the natural interpolations for different interdependent functions are incompletely consistent. For the two-dimensional case [22], this effect is the main source of the error: a similar estimate gives $\delta_{syst} \approx 0.05$, which is larger than the natural summation error for most functions. According to Table 1, central value (17) is well balanced with respect to the first three estimates; for this reason, we do not add $\delta_{syst}$ to its uncertainty and present it in the unchanged form in Fig. 1. An additional argument is that the exponent ν for the two-dimensional case [22] is almost free of systematic errors.

Figure 1 obviously shows that result (17) certainly covers both the central value for the ground experiment [1] and the average value of the satellite experiment [2–4]. The upper bound of Eq. (17) is slightly lower than the most recent experimental value, ν = 0.6709(1) [4], but the difference is at the level of the estimated systematic error given by Eq. (18) and cannot be considered significant. The agreement can be easily ensured by expanding the set of the allowable interpolations or using more complicated interpolation procedures (see, e.g., [22]). Nevertheless, the natural summation results are closer to the lower bound of the experimental values. For example, if only the smoothest interpolations with −0.25 < $\tilde{N}$ < 1.4 are used in Fig. 2b, the result for g* is refined to 1.407 ± 0.001, which gives rise to the decrease in the upper estimate for ν to 0.6703. In contrast, higher estimates proposed for ν in [13–16] seem to have low probabilities, because an unnatural form of the coefficient functions should be assumed in order to obtain those estimates. In view of this circumstance, note (see Table 2) that our value for the exponent $\gamma$ is in complete agreement with the result reported in [17], whereas the most significant discrepancy is associated with the exponent $\eta$; but the summation of the series for the function $\eta(g)$ involves almost no arbitrariness (see Fig. 3b).[6] On the other hand, the Monte Carlo results should not be considered as the results of the direct numerical "measurement," because they are obtained in the course of a complicated and ambiguous treatment [17].

**Table 1.** Various estimate for the exponent ν

| Series | Interval for ν | Central value |
|---|---|---|
| $\nu^{-1}(g)$ | 0.6694 ± 0.6706 | 0.6700 |
| $\gamma^{-1}(g), \eta(g)$ | 0.6693 ± 0.6702 | 0.6698 |
| $\eta_2(g), \eta(g)$ | 0.6698 ± 0.6707 | 0.6702 |
| $\gamma^{-1}(g), \eta_2(g)$ | 0.6654 ± 0.6710 | 0.6682 |

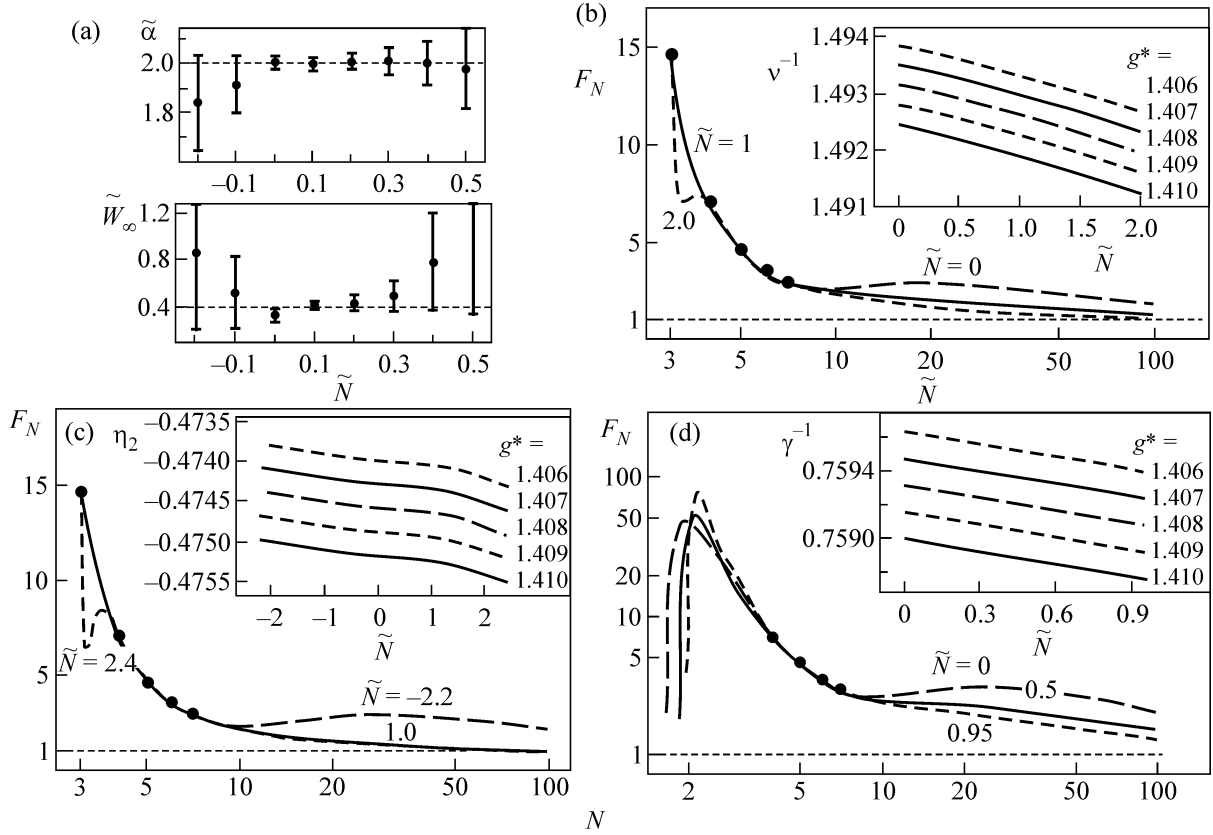

**Fig. 4.** (a) Parameters of the strong-coupling asymptotic expression for the function $\nu^{-1}(g)$ (working interval $20 < N < 40$). (b–d) Interpolation curves for the functions (b) $\nu^{-1}(g)$, (c) $\eta_2(g)$ ($L_0 = 3$), and (d) $\gamma^{-1}(g)$ ($L_0 = 1$). The inset shows the summation results at $g = g*$.

It is easy to see that out results (see Table 2) are in complete agreement with the results of classical works [5–7], but are more accurate because the results of the algorithm used in [5–7] depend strongly on the form of the summation procedure. This uncertainty cannot be properly analyzed, and one use semiempirical rules to restrict it. In our approach, the uncertainty in the results is directly attributed to the ambiguous interpolation of the coefficient functions and, hence, the estimate of their error is completely clear. The strongest refinement occurs for the exponents $\eta$ and $\omega$.

---

[6] Note some conceptual differences in the interpretation of the series for the function $\eta(g)$. The good accuracy in the estimate of the strong-coupling asymptotic expression (see Fig. 3a) shows that the function $\eta(g)$ is a monotonically increasing regular function. This inevitably means that oscillations in its first coefficients rapidly damp and do not continue to the region of large $N$ values. The interpolation for the even and odd $N$ values separately [8] implies that oscillations are damped according to the power law. This gives rise to the exponential increase in the coefficients $U_N$ and Borel image $B(z)$ and is responsible for the singularity in the function $\eta(g)$ at $g \sim 1$.

Let us discuss the correspondence of our results with the results of the variational perturbation theory [8–11], which is a certain interpolation scheme without deep physical meaning. The high accuracy that is sometimes stated in this theory is achieved only due to restriction by a certain interpolation scheme and ignoring its numerous variants. The real uncertainty of the results can be seen by comparing various estimates presented in [11]. In our opinion, higher $\nu$ values obtained in this theory (see Fig. 1 and Table 2) are explained by an inaccurate value of the exponent $\omega$. The formula

$$\nu = 0.6710 + 0.0553(\omega - 0.800)$$

presented in [8, 11] yields $\nu = 0.6698$ for $\omega = 0.778$, which almost coincides with our central value and corresponds to the smoothest interpolation curves.

**Table 2.** Results for the critical exponents in comparison with the results obtained by other authors (numbers in the parentheses are the errors in the units of the last significant digit)

|  | BNM [5] | LG–ZJ [6] | G–ZJ [7] | Kl [8] | CHPV [16] | Present work |
|---|---|---|---|---|---|---|
| $\gamma$ | 1.316(9) | 1.3160(25) | 1.3169(20) | 1.318 | 1.3178(2) | 1.3172(8) |
| $\nu$ | 0.669(3) | 0.6695(20) | 0.6703(15) | 0.6710 | 0.6717(1) | 0.6700(6) |
| $\eta$ | 0.032(15) | 0.033(4) | 0.0354(25) | 0.0356(10) | 0.0381(2) | 0.0334(2) |
| $\eta_2$ | −0.474(8) | −0.4740(25) |  |  |  | −0.4746(9) |
| $\omega$ | 0.780(10) | 0.780(25) | 0.789(11) | 0.800 | 0.785(20) | 0.778(4) |
| $g^*$ | 1.409(5) | 1.406(4) | 1.403(3) |  | 1.4032(7) | 1.408(2) |

It is clear from the above analysis that the results of the natural summation of the series for the renormalization group functions do not indicate the necessity of any systematic shift of the experimental data: their relaxation to a certain average level around which they vary is sufficient. For this reason, we think that the statement made in [16] about unreliability of the experimental data is unjustified.

This work was supported by the Russian Foundation for Basic Research (project no. 06-02-17541).

*Translated by R. Tyapaev*